\documentclass[a4paper]{spie}  

\usepackage{amsmath,amsfonts,amssymb}
\usepackage{graphicx}
\usepackage[colorlinks=true, allcolors=blue]{hyperref}
\usepackage{float}        
\usepackage{placeins}     
\usepackage{booktabs}
\usepackage{caption}
\usepackage{subcaption}
\usepackage{multirow}
\usepackage{natbib} 

\setcitestyle{square,numbers,sort&compress}

\title{Enhancing the Aesthetic Appeal of AI-Generated Physical Product Designs through LoRA Fine-Tuning with Human Feedback}

\author[a]{Dinuo Liao\textsuperscript{*}}
\author[a]{James Derek Lomas}
\author[b]{Cehao Yu}

\affil[a]{Faculty of Industrial Design Engineering, Delft University of Technology, Delft, Netherlands}
\affil[b]{Color, Imaging and Metaverse Research Center, The Hong Kong Polytechnic University, Hong Kong SAR, China}

\authorinfo{D.L. (Corresponding Author): \texttt{liaodinuo98@gmail.com} \\
J.D.L.: \texttt{j.d.lomas@tudelft.nl} \\
C.Y.: \texttt{cehao.yu@polyu.edu.hk}}

\begin{document}
\maketitle

\begin{abstract}
This study explores how Low-Rank Adaptation (LoRA) fine-tuning, guided by human aesthetic evaluations, can enhance the outputs of generative AI models in tangible product design, using lamp design as a case study. By integrating human feedback into the AI model, we aim to improve both the desirability and aesthetic appeal of the generated designs. Comprehensive experiments were conducted, starting with prompt optimization techniques and focusing on LoRA fine-tuning of the Stable Diffusion model. Additionally, methods to convert AI-generated designs into tangible products through 3D realization using 3D printing technologies were investigated. The results indicate that LoRA fine-tuning effectively aligns AI-generated designs with human aesthetic preferences, leading to significant improvements in desirability and aesthetic appeal scores. These findings highlight the potential of human-AI collaboration in tangible product design and provide valuable insights into integrating human feedback into AI design processes.
\end{abstract}


\section{INTRODUCTION}
\label{sec:intro}

Designing tangible products presents unique challenges, especially when integrating artificial intelligence into the design process. Tangible product design requires consideration of aesthetic appeal, functionality, ergonomics, and manufacturability \cite{Norman2004}. \textbf{Aesthetic appeal} refers to the qualities of a design that elicit positive sensory and emotional responses from individuals \cite{Hekkert2008}. Understanding and integrating human aesthetic preferences is crucial to create products that resonate with users.

Generative AI has revolutionized creative industries by automating aspects of design and ideation \cite{Goodfellow2014}. Models like Generative Adversarial Networks (GANs) and diffusion models have been instrumental in producing high-quality, novel content \cite{Karras2019,Dhariwal2021,Rombach2022}. In industrial design, these models can rapidly generate diverse product concepts, aiding designers in exploring a vast design space \cite{Burnap2019}.

Despite these advancements, aligning AI-generated outputs with human aesthetic preferences remains challenging. Traditional fine-tuning methods may not effectively capture nuanced human feedback, making it difficult to tailor AI outputs to meet specific aesthetic criteria \cite{Ziegler2019,Howard2018}. Our experiments relate to empirical aesthetics by integrating human evaluations directly into the AI model, aiming to enhance the aesthetic appeal of generated designs \cite{Crilly2004,Hekkert2008}.

\begin{figure}
\centering
\includegraphics[width=1.0\textwidth]{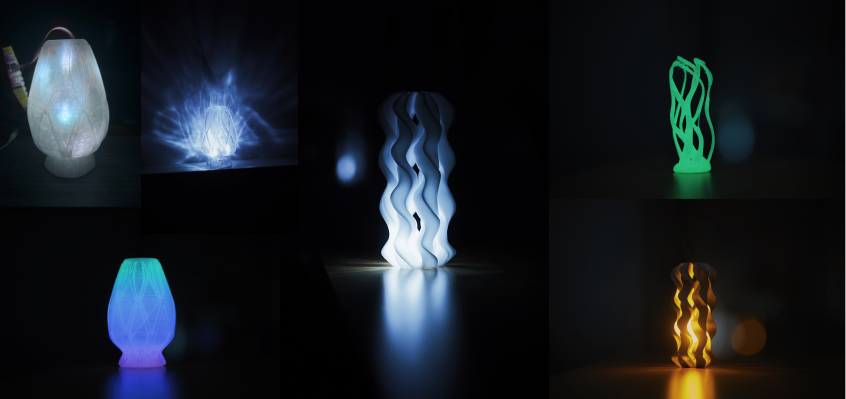}
\caption{Some examples of 3D-printed lamps produced by the generative AI design process}
\label{fig:printed_lamp}
\end{figure}

To illustrate the types of outcomes we aim to achieve, we present several 3D-printed lamp designs in Figure \ref{fig:printed_lamp}. These examples highlight the physical products that can be realized from AI-generated concepts. Building upon previous work on prompt optimization, this research focuses on applying LoRA fine-tuning, guided by human aesthetic evaluations, to enhance AI-generated designs in tangible product design. We utilized different AI models in a novel manner to achieve results consistent with human perception. Additionally, we explored methods to convert AI-generated designs into tangible products through 3D realization.

\section{METHODS}
\label{sec:methods}

\subsection{Participants and Ethics}
Participants from Delft University of Technology volunteered without financial compensation. All reported normal or corrected-to-normal visual acuity and normal color vision, confirmed using the Ishihara test under standardized lighting. The study adhered to ethical standards outlined in the World Medical Association’s Declaration of Helsinki, Dutch law, and local institutional guidelines. Ethical approval was granted by the Human Research Ethics Committee (HREC) of Delft University of Technology, and written informed consent was obtained from all participants.

\subsection{Materials and Tools}
We used the Stable Diffusion model \cite{Rombach2022} for generating lamp designs. The LoRA framework was implemented using PyTorch for efficient adaptation \cite{Hu2021}. A custom web application collected human evaluations. For 3D realization, AI-generated 3D model tools created mesh models, and Cinema 4D (C4D) refined them for 3D printing. Fused Deposition Modeling (FDM), Stereolithography (SLA), and Selective Laser Sintering (SLS) produced physical prototypes. Data analysis was performed using Python and Matplotlib.

\subsection{Procedure}

\paragraph{Initial Experiment.}
We assessed the impact of prompt optimization on AI-generated lamp designs. Participants provided prompts optimized using a custom GPT-4 model. Designs generated from original and optimized prompts were evaluated on three attributes using a scale from 0 to 100: \textbf{Desirability (Initial Experiment)}, \textbf{Printability}, and \textbf{Alignment} with the original concept.

\paragraph{Main Experiment with LoRA Fine-Tuning.}
The main experiment focused on fine-tuning the Stable Diffusion model using LoRA to align with human aesthetic preferences. In this context, \textbf{Desirability (LoRA Fine-Tuning)} refers to the overall attractiveness and appeal of the design from a user's perspective, considering functionality, usability, and emotional response. \textbf{Aesthetic appeal} focuses on the visual attractiveness of the design, emphasizing elements like form, style, and color.

From approximately 1,600 generated lamp designs, we selected 405 high-quality images. Each image was rated using a ``like'' or ``dislike'' scale for \textbf{Desirability (LoRA Fine-Tuning)} and \textbf{Aesthetic appeal}, receiving 5 to 15 ratings to ensure reliability. The \textbf{like rate} for each image was calculated using:
\begin{equation}
\text{Like Rate} = \left( \frac{\text{Number of ``like'' responses}}{\text{Total responses}} \right) \times 100\%.
\end{equation}

Images with like rates greater than 70\% were considered as having high \textbf{Desirability (LoRA Fine-Tuning)} or \textbf{Aesthetic appeal}. We selected these high-rated images for training: 40 for desirability and 34 for aesthetic appeal. These images and their prompts formed the training datasets for two separate LoRA fine-tuning processes.

The LoRA technique \cite{Hu2021} was implemented with the following parameters: batch size of 4, learning rate of 0.0001, and 10 epochs. Mixed precision training (\texttt{fp16}) optimized memory usage and training speed. The LoRA weight was set to 0.6 after empirical testing. Model outputs were generated at 512x512 pixels using the Euler a sampling method with 20 steps.

After fine-tuning, 100 images for each dimension were generated. Each image was rated 25 to 35 times using the same ``like'' or ``dislike'' scale. The like rates before and after fine-tuning were compared to assess the effectiveness of the LoRA technique.

\paragraph{3D Realization.}
We selected high-rated designs to convert into tangible products. AI-generated 3D model tools created mesh models from 2D images, and C4D refined them for 3D printing. Physical prototypes were produced using FDM, SLA, and SLS printing methods, including an aluminum lamp using SLS. These prototypes allowed us to assess manufacturability and identify challenges in translating digital designs into physical objects.

\paragraph{Data Analysis.}
Data analysis was performed using Python. To assess the effects of LoRA fine-tuning on like rates, we employed both parametric and non-parametric statistical tests based on the data distribution. For normally distributed data, paired t-tests were used; for nonnormally distributed data, Mann-Whitney U tests were employed. Statistical significance was evaluated with p-values.

\section{FINDINGS}
\label{sec:findings}

\subsection{Initial Experiment Results}
Prompt optimization significantly improved \textbf{Desirability (Initial Experiment)} (p=$3.37 \times 10^{-5}$) from mean 58.4 (SD=12.5) to 67.2 (SD=13.1). No improvement in \textbf{Printability} (p=0.17) or \textbf{Alignment} (p=0.20) was observed. Table \ref{tab:overall-quantitative-results} summarizes these results.

\subsection{Main Experiment Results with LoRA Fine-Tuning}

Table \ref{tab:overall-quantitative-results} summarizes the results of both the initial experiment and the main experiment. LoRA fine-tuning demonstrated significant improvements in both desirability and aesthetic appeal.

\begin{figure}
\centering
\begin{subfigure}{0.43\textwidth}
    \centering
    \includegraphics[width=\textwidth]{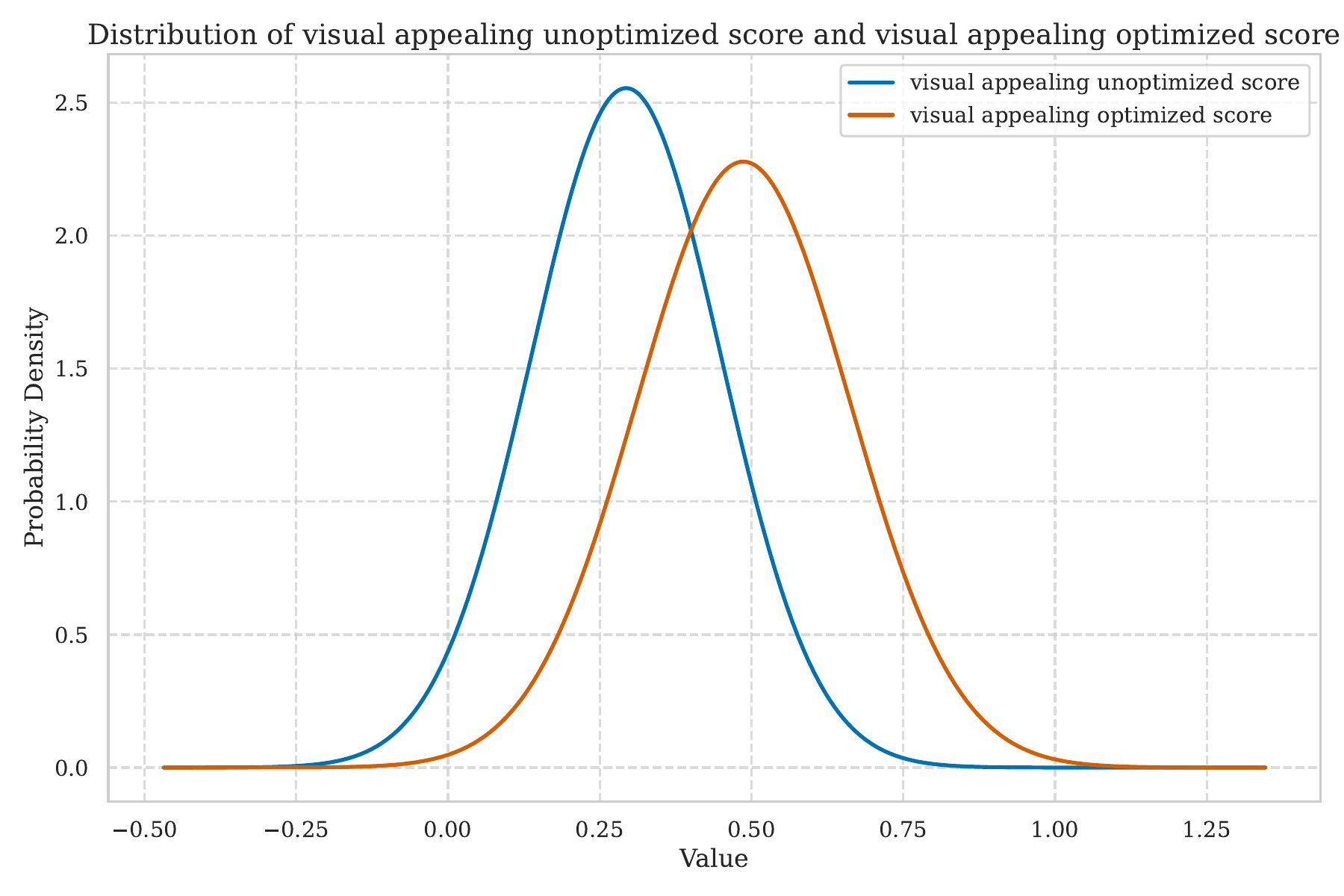}
    \caption{Aesthetic Appeal Score Distribution}
    \label{fig:aesthetic_histogram}
\end{subfigure}
\hfill
\begin{subfigure}{0.46\textwidth}
    \centering
    \includegraphics[width=\textwidth]{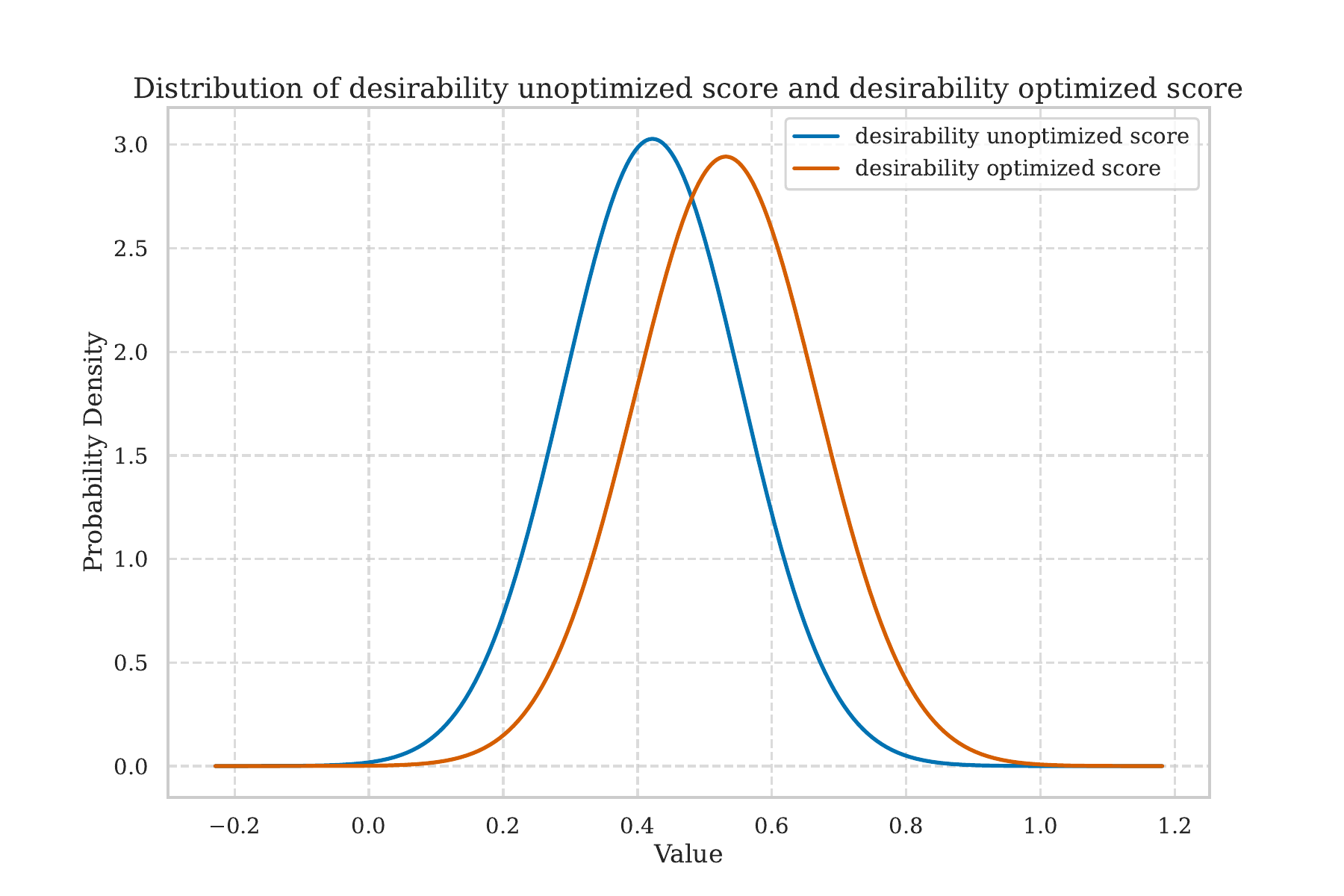}
    \caption{Desirability Score Distribution}
    \label{fig:desirability_histogram}
\end{subfigure}
\caption{Score Distributions for Aesthetic Appeal and Desirability Before and After LoRA Fine-Tuning}
\label{fig:combined_histograms}
\end{figure}

\paragraph{Desirability (LoRA Fine-Tuning)}
After LoRA fine-tuning, the mean like rate for \textbf{Desirability (LoRA Fine-Tuning)} increased from 63.5\% (SD = 6.0) to 81.2\% (SD = 5.2), with a p-value of 0.0001, indicating a highly significant improvement. Figure \ref{fig:desirability_histogram} shows the distribution of desirability scores before and after fine-tuning.

\paragraph{Aesthetic Appeal}
Aesthetic appeal like rate increased from 61.8\% (SD = 6.3) to 84.7\% (SD = 5.0), p=$9.8 \times 10^{-8}$. Figure \ref{fig:aesthetic_histogram} illustrates the distribution of aesthetic appeal scores.

These results, summarized in Table \ref{tab:overall-quantitative-results}, highlight the effectiveness of LoRA fine-tuning in aligning AI-generated designs with human preferences.

\begin{figure}
\centering
\begin{subfigure}{0.48\textwidth}
    \centering
    \includegraphics[height=4cm]{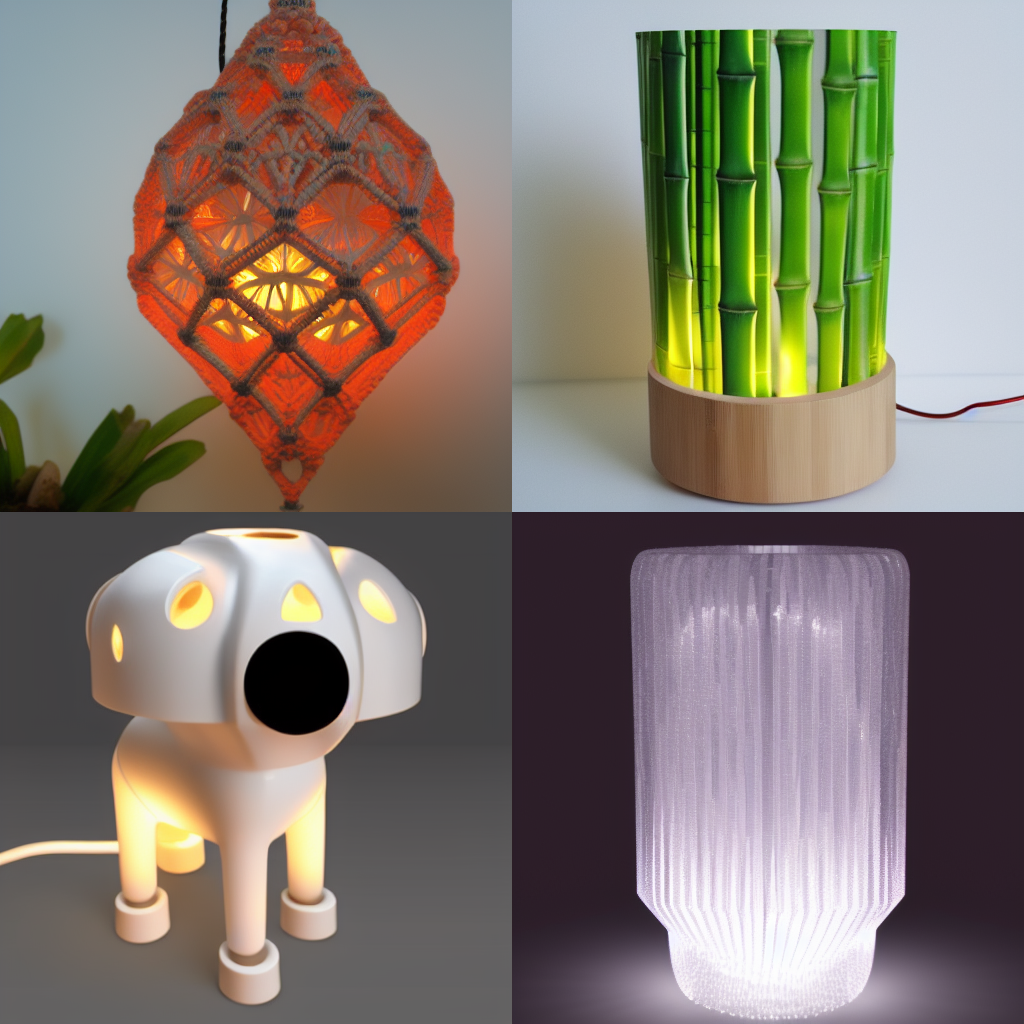}
    \caption{Before LoRA Fine-Tuning (Desirability)}
    \label{fig:desirability_before}
\end{subfigure}
\hfill
\begin{subfigure}{0.48\textwidth}
    \centering
    \includegraphics[height=4cm]{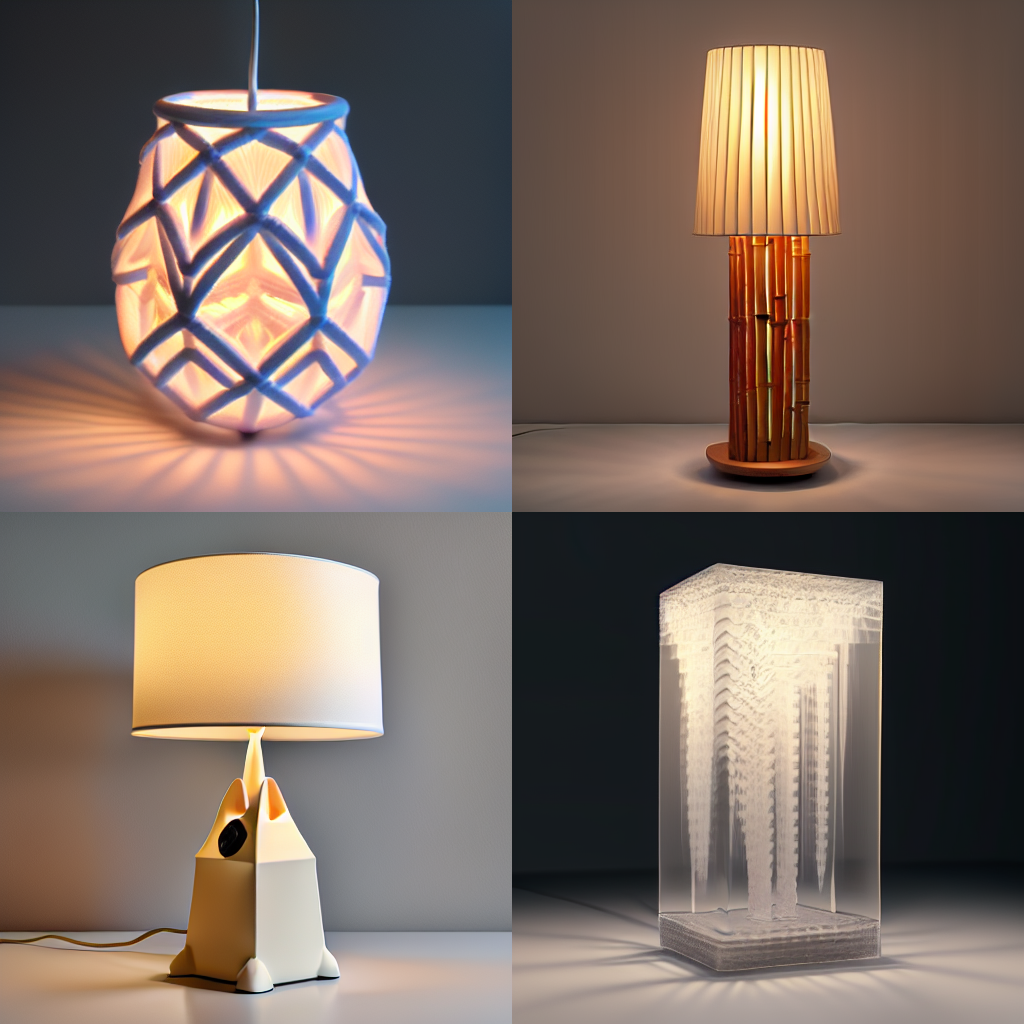}
    \caption{After LoRA Fine-Tuning (Desirability)}
    \label{fig:desirability_after}
\end{subfigure}
\vspace{5mm}
\begin{subfigure}{0.48\textwidth}
    \centering
    \includegraphics[height=4cm]{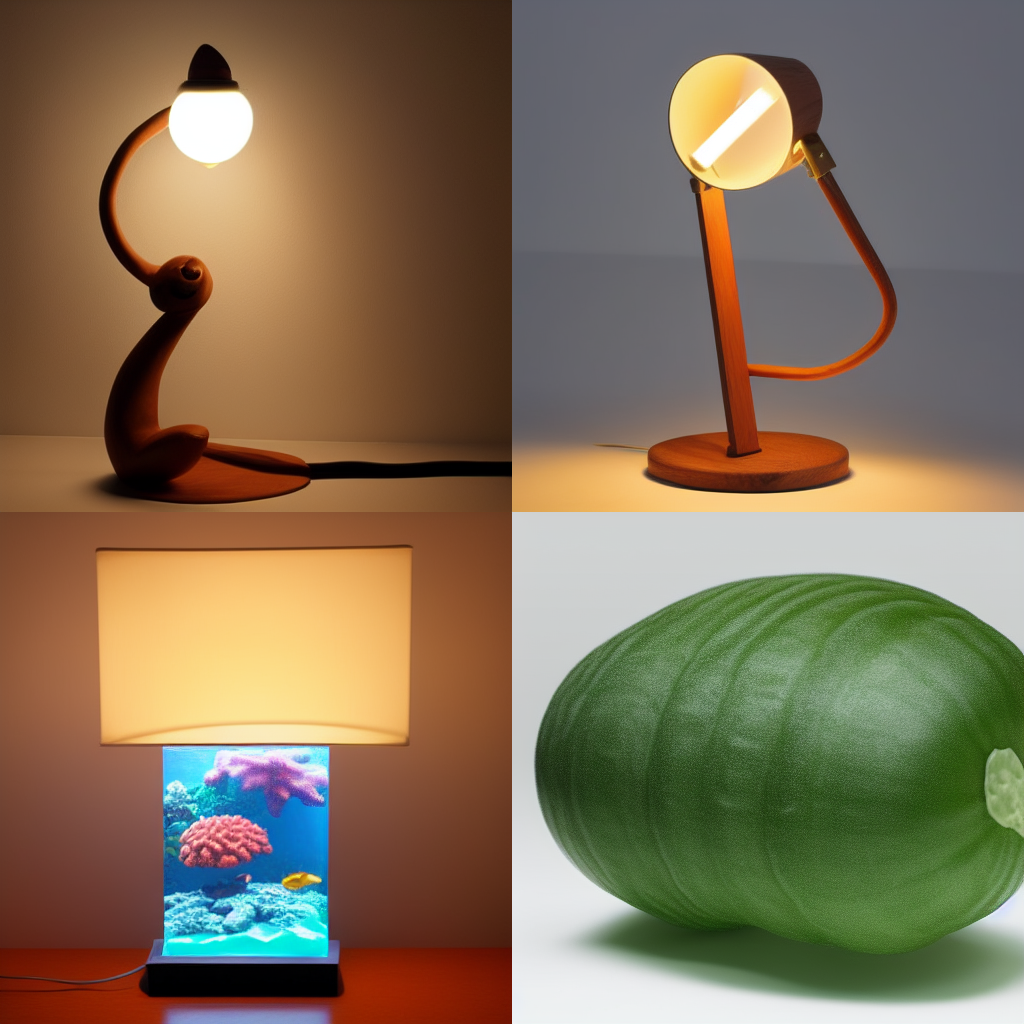}
    \caption{Before LoRA Fine-Tuning (Aesthetic Appeal)}
    \label{fig:aesthetic_before}
\end{subfigure}
\hfill
\begin{subfigure}{0.48\textwidth}
    \centering
    \includegraphics[height=4cm]{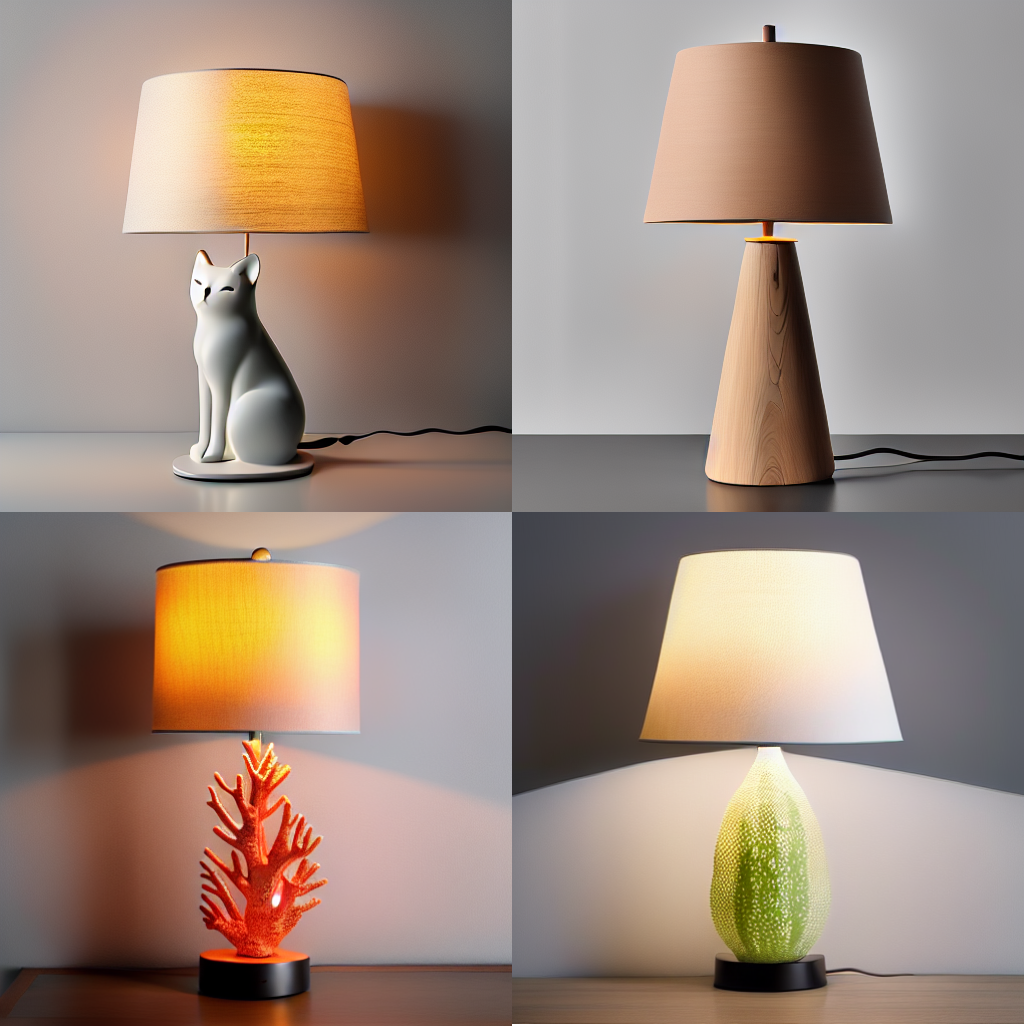}
    \caption{After LoRA Fine-Tuning (Aesthetic Appeal)}
    \label{fig:aesthetic_after}
\end{subfigure}
\caption{Comparison of Lamp Designs Before and After LoRA Fine-Tuning for Desirability and Aesthetic Appeal}
\label{fig:combined_design_images_corrected}
\end{figure}

\begin{table}
\centering
\begin{tabular}{|l|l|l|l|l|}
\hline
\textbf{Metrics} & \textbf{Experiments}       & \textbf{Unoptimized Mean} & \textbf{Optimized Mean} & \textbf{p-value} \\ \hline
Desirability     & \multirow{2}{*}{LoRA Finetuned} & 63.5 (6.0)               & 81.2 (5.2)              & $10^{-4}$        \\ \cline{1-1} \cline{3-5}
Aesthetic Appeal &                               & 61.8 (6.3)               & 84.7 (5.0)              & $9.8 \times 10^{-8}$     \\ \hline
Desirability     & \multirow{3}{*}{Initial Experiments} & 58.4 (12.5)             & 67.2 (13.1)             & $3.37 \times 10^{-5}$    \\ \cline{1-1} \cline{3-5}
Printability     &                              & 62.1 (11.8)              & 63.7 (12.0)             & 0.17             \\ \cline{1-1} \cline{3-5}
Alignment        &                              & 59.8 (10.9)              & 61.0 (11.2)             & 0.20             \\ \hline
\end{tabular}
\caption{Overall Quantitative Results of Initial Experiment and Main Experiment (LoRA Fine-Tuning).}
\label{tab:overall-quantitative-results}
\end{table}

Figure \ref{fig:combined_design_images_corrected} shows example lamp designs before and after LoRA fine-tuning.

\subsection{3D Realization}
High-rated designs were transformed into physical prototypes. AI-generated 3D model tools created mesh models from 2D images, and C4D refined them for 3D printing. Models were printed using FDM, SLA, and SLS (including aluminum), demonstrating durable and functional products. This process assessed manufacturability and highlighted challenges in translating digital designs into physical objects.

\section{DISCUSSION}
\label{sec:discussion}

Integrating human feedback into AI models through LoRA fine-tuning significantly enhances alignment with human aesthetic preferences. The improvements in \textbf{Desirability} and \textbf{Aesthetic appeal} highlight this approach’s effectiveness. By employing LoRA as a preference model, we efficiently incorporated empirical aesthetic evaluations, addressing limitations of traditional fine-tuning methods \cite{Ziegler2019,Hu2021}.

The successful 3D realization underscores practical applicability. Utilizing AI-generated 3D model tools and C4D bridged the gap between digital designs and tangible products without relying on traditional CAD software, streamlining the transition from concept to prototype.

\paragraph{Implications for Tangible Product Design}
\label{sec:implications}
These findings have significant implications. Designers can leverage LoRA fine-tuning to produce AI-generated designs that closely align with user preferences, resulting in functional and aesthetically pleasing products. The method reduces iterative cycles, saves time and resources, and democratizes advanced generative models for designers without extensive AI expertise. Converting AI-generated designs into physical products proves feasibility and enhances the design pipeline from ideation to manufacturing.

This approach fosters human-AI collaboration, enabling innovative products that better meet consumer demands. LoRA fine-tuning is also scalable, adapting rapidly to various aesthetics across multiple product categories.

\paragraph{Limitations and Future Work}
\label{sec:limitations}
While focused on lamp design, future studies could test other product categories. Investigating automated 3D modeling from 2D images and integrating more diverse forms of human feedback (e.g., qualitative reviews) could provide deeper insights. Exploring the long-term impact of AI-assisted design on creativity and designer roles would further clarify the evolving landscape of product design industries.

Future work could also explore the long-term impacts of AI-assisted design on the creative process, including how it influences designers’ roles and the overall innovation landscape in product design industries.

\section{CONCLUSION}
\label{sec:conclusion}

Integrating human ratings into AI models using LoRA significantly enhances \textbf{Desirability} and \textbf{Aesthetic appeal} in tangible product design. Incorporating human preferences and exploring 3D realization improves design quality, highlighting the potential of human-AI collaboration for more efficient, consumer-aligned design.

Moreover, transforming AI-generated designs into physical prototypes validates practical viability, enabling rapid prototyping and iterative design. Human feedback ensures products resonate with user preferences, enhancing marketability and satisfaction.

These insights extend beyond lamp design, suggesting applicability across various product categories to foster creativity, efficiency, and consumer alignment. As AI evolves, human creativity and machine intelligence synergy will shape the future of product design, advancing both aesthetics and functionality.

\acknowledgments 
We sincerely thank all participants for their valuable contributions. We also acknowledge the support of the Faculty of Industrial Design Engineering at Delft University of Technology.

\bibliographystyle{spiebib}
\bibliography{report}

\end{document}